# Title: Expansion of a Valley-Polarized Exciton Cloud in a 2D Heterostructure

**Authors:** Pasqual Rivera[1,†], Kyle L. Seyler[1,†], Hongyi Yu[2], John R. Schaibley[1], Jiaqiang Yan[3,4], David G. Mandrus[3,4,5], Wang Yao[2], Xiaodong Xu[1,6]

**Affiliations:**

[1]Department of Physics, University of Washington, Seattle, Washington, 98195, USA

[2]Department of Physics and Center of Theoretical and Computational Physics, University of Hong Kong, Hong Kong, China

[3]Materials Science and Technology Division, Oak Ridge National Laboratory, Oak Ridge, Tennessee, 37831, USA

[4]Department of Materials Science and Engineering, University of Tennessee, Knoxville, Tennessee, 37996, USA

[5]Department of Physics and Astronomy, University of Tennessee, Knoxville, Tennessee, 37996, USA

[6]Department of Materials Science and Engineering, University of Washington, Seattle, Washington, 98195, USA

*Correspondence to: xuxd@uw.edu.

†These authors contributed equally to the work.

**Abstract**:

Heterostructures comprising different monolayer semiconductors provide a new system for fundamental science and device technologies, such as in the emerging field of valleytronics. Here, we realize valley-specific interlayer excitons in monolayer $WSe_2$-$MoSe_2$ vertical heterostructures. We create interlayer exciton spin-valley polarization by circularly polarized optical pumping and determine a valley lifetime of 40 nanoseconds. This long-lived polarization enables the visualization of the expansion of a valley-polarized exciton cloud over several microns. The spatial pattern of the polarization evolves into a ring with increasing exciton density, a manifestation of valley exciton exchange interactions. Our work lays a foundation for quantum optoelectronics and valleytronics based on interlayer excitons in van der Waals heterostructures.

**Main Text:**

Van der Waals heterostructures of two-dimensional (2D) materials provide an exciting platform for engineering artificial material systems with unique properties (*1*). A beautiful example is the demonstration of the Hofstadter butterfly physics in moiré superlattice structures composed of graphene and hexagonal boron nitride (*2-4*). As the library of 2D crystals is explored further, the range of possible new phenomena in condensed matter physics becomes ever more diverse. For example, heterostructures of 2D semiconductors (namely, transition metal dichalcogenide monolayers, $MX_2$) have been assembled with type-II band alignment (*5-8*), where electrons and holes energetically favor different layers (Fig. 1A). These heterostructures form atomically thin p-n junctions which can be used for photon-energy harvesting (*9-15*), and host interlayer excitons ($X_I$) with the Coulomb-bound electron and hole located in different monolayers (*14-17*), analogous to the indirect excitons in coupled III-V quantum wells (*18, 19*). This species of exciton has a lifetime far exceeding those in monolayer $MX_2$, and the vertical separation of holes and electrons entails a permanent out-of-plane electric dipole moment, providing an optical means to pump interlayer electric polarization, and facilitating electrical control of interlayer excitons (*17*).

Distinct from coupled quantum wells, $MX_2$ heterostructures possess several unique features. First, the monolayers' band edges are at doubly degenerate corners (rather than the center) of the hexagonal Brillouin zone, so $X_I$ has an internal degree of freedom specified by the combination of electron and hole valley indices (*20*). Second, the twist angle between the crystal axes of constituent monolayers can be selected to control the optoelectronic properties (*21-23*), such as the dipole strength and interlayer exciton lifetime (*24*). Third, the constituent $MX_2$ monolayers exhibit unique valley-contrasting physical properties, such as spin-valley locking (i.e. each valley has a single and distinct spin species), optical selection rules (*25-27*), and Berry curvature (*20*). The inheritance of valley physics in twisted $MX_2$ heterostructures gives rise to unprecedented optical and transport properties of $X_I$ (*24*), allowing the possibility of excitonic optoelectronic circuits with valley functionalities and a new platform for investigating excitonic superfluidity and condensation (*28*).

In this work, we observe long valley lifetime and long-distance valley drift-diffusion of $X_I$ in $MoSe_2$-$WSe_2$ heterostructures with small twist angles. Using circularly polarized optical pumping, we demonstrate interlayer exciton valley polarization with electrostatic control. Time-resolved photoluminescence (PL) measurements reveal valley polarization lifetimes of 40 ns, orders of magnitude longer than in monolayers (*29-31*). Remarkably, we observe valley-polarized PL from a region substantially larger than the excitation profile. The spatial pattern of polarization evolves into a ring under increasing excitation intensity, demonstrating a difference in the drift of the majority and minority valley exciton species, a consequence of repulsive valley exciton exchange interactions.

Our devices consist of a pair of exfoliated monolayers of $WSe_2$ and $MoSe_2$, which are stacked by dry transfer (*32*) on a 285 nm insulating layer of $SiO_2$ on a silicon substrate. Standard electron beam lithography techniques are used to fabricate metallic contacts (V/Au) to the heterostructure, and the silicon substrate functions as a global backgate (*33*). The optical brightness of the $X_I$ depends sensitively on the relative alignment of the two constituent monolayers. Theory shows that for twist angle near zero or 60°, there exist light cones at small kinematic momenta where the $X_I$ can directly interconvert with photons (*24*). In such heterostructures, the $X_I$ can radiatively recombine after scattering into these light cones, e.g. by exciton-phonon or exciton-exciton interactions (*24*). As such, the $X_I$ can be observed in PL, and in our study, we focus on this type of heterostructure. To fabricate such samples, we identify the armchair axes of individual monolayers by polarization-resolved second-harmonic generation and then align these axes (Fig. S1). This yields heterostructures with twist angles near zero or 60° (*33*). All data in the main text are taken at 30 K from the device shown in the optical microscope image in Fig. 1B, with $WSe_2$ stacked on $MoSe_2$, and the excitation laser energy in resonance with the A exciton of $WSe_2$ (1.72 eV).

We first perform polarization-resolved PL at zero gate voltage ($V_g$ = 0 V). We apply circularly polarized continuous wave laser excitation and separately detect the right circular ($\sigma^+$) and left circular ($\sigma^-$) PL. Figure 1C shows the $\sigma^+$ (black) and $\sigma^-$ (red) components of the $X_I$ PL under circularly polarized excitation. These results show that $X_I$ emission is strongly co-polarized with the incident light. Denoting the degree of polarization by $\rho = \frac{I_+ - I_-}{I_+ + I_-}$, where $I_\pm$ is the intensity of the $\sigma_\pm$ PL components, we observe $|\rho_{max}| > 0.3$. Similar results are obtained from several other samples (Fig. S2).

The observation of polarized PL demonstrates that the interlayer excitons can retain memory of the excitation light helicity, which is a consequence of the valley optical selection rules in 2D heterostructures (*24*). In the following, we discuss the generation of valley polarization in heterostructures with near AA-like stacking (twist angle near 0°, illustrated in Fig. 1D), but similar conclusions can be drawn for heterostructures near AB-like stacking (twist angle near 60°, Fig. S3) (*33*). The valley configuration of an interlayer exciton is specified by the valley indices of its electron and hole. With the spin-valley locking in monolayer $MX_2$, a universal assignment of the valley index is applicable in the twisted heterostructures, and herein we denote the valley with electron spin up (down) as $+K$ ($-K$) in both layers. First, $\sigma^+$ excitation creates valley-polarized intralayer excitons in the $+K_W$ valley in $WSe_2$ and $+K_M$ valley in $MoSe_2$. Next, charge carriers relax to the heterostructure band edges through interlayer charge transfer on sub-picosecond time scales (*11, 15*) to form $X_I$. Due to the large momentum difference, interlayer hopping between $+K_W$ and $-K_M$ valleys is strongly suppressed. Conversely, the $+K_W$ and $+K_M$ valleys have small momentum mismatch and the spin-conserving interlayer hopping between these valleys becomes the dominant relaxation channel. Therefore, the $\sigma^+$ excitation leads to valley-polarized $X_I$, as illustrated in Fig. 1E. The situation for $\sigma^-$ excitation can be obtained by time reversal. The radiative recombination of the valley-polarized $X_I$ is facilitated by the interlayer coupling, which allows emission of photons that are co-polarized with the excitation source (*24*).

We find that the degree of $X_I$ valley polarization can be electrically controlled by the gate. Figure 2A shows polarization-resolved PL spectra at selected $V_g$ under $\sigma^+$ excitation with ~50 ps laser pulses. There is a strong gate dependence of the valley polarization, which is greatest at +60 V and highly suppressed at -60 V (see Fig. S4 for the full data set). Time-resolved measurements reveal the dynamics of the $X_I$ PL polarization. In Fig. 2B, we show the decay of co-polarized (black) and cross-polarized (red) interlayer exciton PL, as well as the degree of polarization (blue), at the same $V_g$ values as in Fig. 2A (see Fig. S5 for the full data set). The valley polarization lifetime increases with $V_g$, reaching $39\pm2$ ns at +60 V, as determined by fitting a single exponential decay. Long valley lifetimes were also measured in heterostructures with the opposite stacking order (i.e. MoSe$_2$ on WSe$_2$, Fig. S6). These measurements imply a strong suppression of intervalley scattering for interlayer excitons and a valley lifetime several orders of magnitude longer than that of intralayer excitons in monolayers, where valley depolarization occurs on picosecond timescales (*29-31*).

The long valley lifetime of the X$_I$ allows visualization of their lateral drift and diffusion. Figure 3A displays a sequence of spatial maps of the X$_I$ PL polarization under pulsed excitation (40 MHz repetition rate) at $V_g = 60$ V for selected average excitation powers. The spatial pattern of $\rho$ shows the evolution of a ring with diameter that increases with excitation intensity (see full data set in Fig. S7). The pattern of polarization stands in stark contrast to the spatial distribution of the emission. Figure 3B shows the polarization-resolved PL intensity spatial maps at 20 μW, where both $\sigma^+$ and $\sigma^-$ PL components display an approximately Gaussian profile centered at the excitation spot. For direct comparison of the different spatial profiles, Fig. 3C gives the average PL intensity and $\rho$ as a function of the distance from the beam center for the 40 μW case. The data shows the drift-diffusion of $\sigma^+$ (black) and $\sigma^-$ (red) polarized excitons away from the laser spot (0.7 μm FWHM, dashed) as well as the ring of larger $\rho$ (blue), demonstrating the striking difference between the spatial distribution of polarization and the total density of X$_I$.

One possible explanation for the observed polarization ring is density-dependent intervalley scattering. However, consideration of the valley polarization as a function of excitation intensity suggests otherwise. Fig. 3D shows the power dependence of valley polarization from the spatially integrated $I_\pm$, which decreases by about 25% over a 60-fold increase in excitation power (corresponding to at least an order of magnitude increase in the exciton density). Yet, the radial dependence of $\rho$ shows a decrease of nearly 40% from its peak, at the radius of the ring of polarization, to its minimum at the excitation center (Fig. 3C). This is a dramatic change in $\rho$ considering that the exciton density drops by only a factor of two over this region (i.e., the corresponding PL intensity decreases by approximately 50%). The weak dependence of the integrated valley polarization on excitation intensity, contrasted with the strong variation of $\rho$ in the ring feature, implies that density-dependent intervalley scattering is not the dominant mechanism in the ring formation.

Rather, the observed spatial patterns in the valley polarization can be understood well as manifestations of valley-dependent many-body interactions in the dense interlayer exciton gas (*33*). The spin-valley polarized X$_I$, which possess out-of-plane dipoles, interact through dipole-dipole and exchange interactions, both of which are repulsive. Due to the small interlayer separation of about 7 Å, we estimate that the short-ranged exchange interaction is stronger than the dipole-dipole repulsion (*33*). Since the exchange interactions are appreciable only between excitons of the same valley species (*33*), in a cloud of valley-polarized interlayer excitons, the majority valley excitons experience stronger mutually repulsive force (Fig. 4A), leading to more rapid expansion than the minority valley excitons (Fig. 4B). On the other hand, the density gradient of excitons will also give rise to diffusion, which is valley-independent and does not produce a ring pattern. Therefore, the relative strength of the diffusion and valley-dependent drift controls the pattern of the spatial polarization. If the interlayer exciton density is large enough that the valley-dependent repulsive interaction dominates the expansion of the exciton gas, higher valley polarization can appear away from the excitation center (Fig. 4B). Indeed, a pronounced ring in the polarization is generated at sufficiently high excitation intensity, as seen in Fig. 3A.

We simulate the $X_I$ expansion using a phenomenological model which accounts for both diffusion and drift due to the valley-dependent repulsive interactions (*33, 34*). The time-dependent spatial density of $\sigma^\pm$ interlayer valley excitons is denoted by $N_\pm(r,t)$ and the model takes the form

$$\frac{\partial N_\pm}{\partial t} = G_\pm + D\nabla^2 N_\pm + \nabla \cdot \left(N_\pm(\alpha_0 \nabla N_\pm + \beta_0 \nabla N_\mp)\right) - \frac{N_\pm}{\tau} - \frac{N_\pm - N_\mp}{\tau_v}.$$

Here, $G_+$ and $G_-$ are the rates of $X_I$ formation co- and cross-polarized with the optical pump, respectively, $D$ is the diffusion coefficient, and $\tau$ and $\tau_v$ are the lifetimes of interlayer exciton and valley polarization, respectively. The third term on the right captures the valley-dependent repulsive interactions, in which the density gradient acts as a driving force for the drift. The force between the excitons in the same valley (the $\alpha_0$ part) is the sum of dipole and exchange interactions, while the force between excitons in different valleys (the $\beta_0$ part) is the dipole interaction only. This simple model captures the essential experimental features, with the simulated power dependence of the spatial polarization (Fig. 4C) showing the formation and evolution of a ring pattern (see Fig. S10 for simulation details).

The ability demonstrated here to generate and control $X_I$ valley polarization with long lifetimes and long-range valley transport presents new opportunities for quantum optoelectronics with valley functionalities, such as spin and valley Hall effects, investigation of many-body interaction physics, and the study of Berry phase effects on exciton quantum dynamics (*24*).

34. J. R. Leonard, Y. Y. Kuznetsova, S. Yang, L. V. Butov, T. Ostatnický, A. Kavokin, A. C. Gossard, Spin transport of excitons. *Nano Lett.* **9**, 4204-4208 (2009).

**Acknowledgments:** We thank David Cobden and Feng Wang for helpful discussion. This work is mainly supported by the Department of Energy, Basic Energy Sciences, Materials Sciences and Engineering Division (DE-SC0008145 and SC0012509). The spectroscopy work is partially supported by NSF-EFRI-1433496. HY and WY were supported by the Croucher Foundation (Croucher Innovation Award), and the RGC and UGC of Hong Kong (HKU17305914P, HKU9/CRF/13G, AoE/P-04/08). JY and DM were supported by US DoE, BES, Materials Sciences and Engineering Division. XX acknowledges a Cottrell Scholar Award and support from the State of Washington funded Clean Energy Institute. Device fabrication was performed at the University of Washington Microfabrication Facility and NSF-funded Nanotech User Facility.

**Additional Author notes:** XX and WY conceived and supervised the project. PR and KS fabricated the samples and performed the experiments, assisted by JS. PR, KS, and XX analyzed data. HY and WY provided theoretical support and performed the simulation. JY and DM synthesized and characterized the bulk crystal. PR, KS, XX, WY, and HY co-wrote the paper. All authors discussed the results.

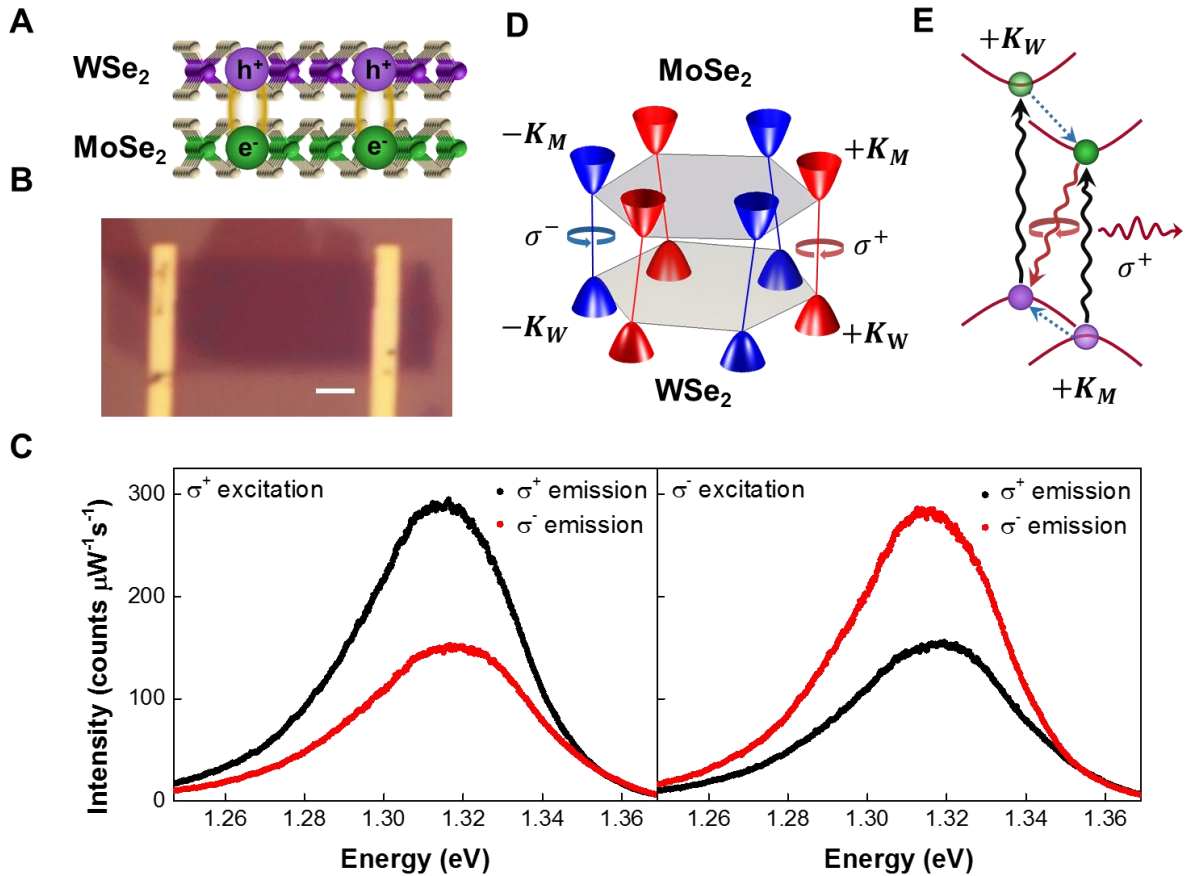

**Figure 1 | Interlayer exciton spin-valley polarization in MoSe$_2$-WSe$_2$ heterostructures.**
**(A)** Side view of MoSe$_2$-WSe$_2$ crystal depicting the interlayer exciton, with holes (h$^+$) and electrons (e$^-$) located in WSe$_2$ and MoSe$_2$, respectively. **(B)** Optical image of device with WSe$_2$ on top of MoSe$_2$. Scale bar is 2 µm. **(C)** Circular polarization-resolved photoluminescence spectra of the interlayer exciton showing the generation of strong valley polarization. **(D)** Illustration of the Dirac points in the hexagonal Brillouin zone of a MoSe$_2$-WSe$_2$ heterostructure with small twisting angle. The $+K$ (red) and $-K$ (blue) valleys at the conduction band minimum (in MoSe$_2$), and valence band maximum (in WSe$_2$) are nearly aligned in momentum space. **(E)** Schematic of the interlayer exciton in the $+K$ valley. First, $\sigma^+$ circularly polarized light (black wavy lines) excites intralayer excitons in the $+K_M$ and $+K_W$ valleys. Fast interlayer charge hopping (blue dotted lines) forms the interlayer exciton in the $+K$ valley. The optical selection rules in the $+K_W$ and $+K_M$ valleys produce co-polarized photoluminescence.

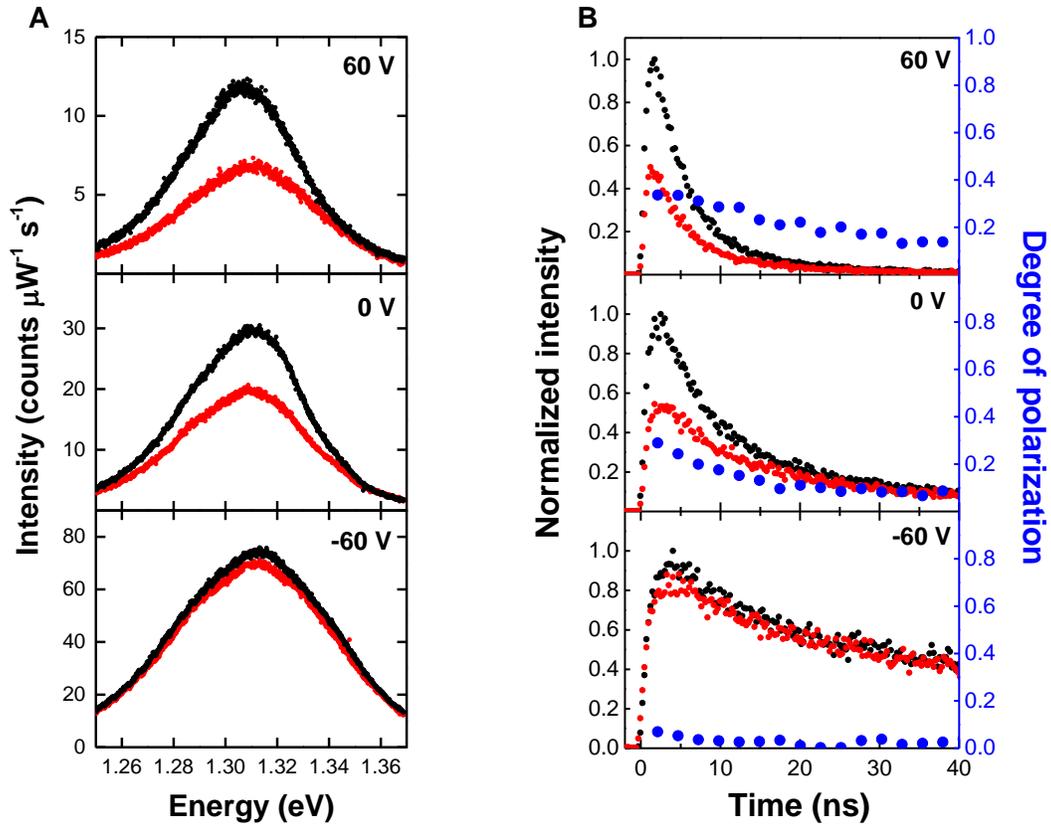

**Figure 2 | Gate-tunable interlayer exciton valley polarization and lifetime.** All plots are for $\sigma^+$ pulsed laser excitation with co-polarized and cross-polarized photoluminescence shown in black and red, respectively. (**A**) Polarization-resolved interlayer exciton photoluminescence at selected gate voltages. (**B**) Time-resolved interlayer exciton photoluminescence at selected gate voltages. The blue curve (right axis) shows the decay of valley polarization.

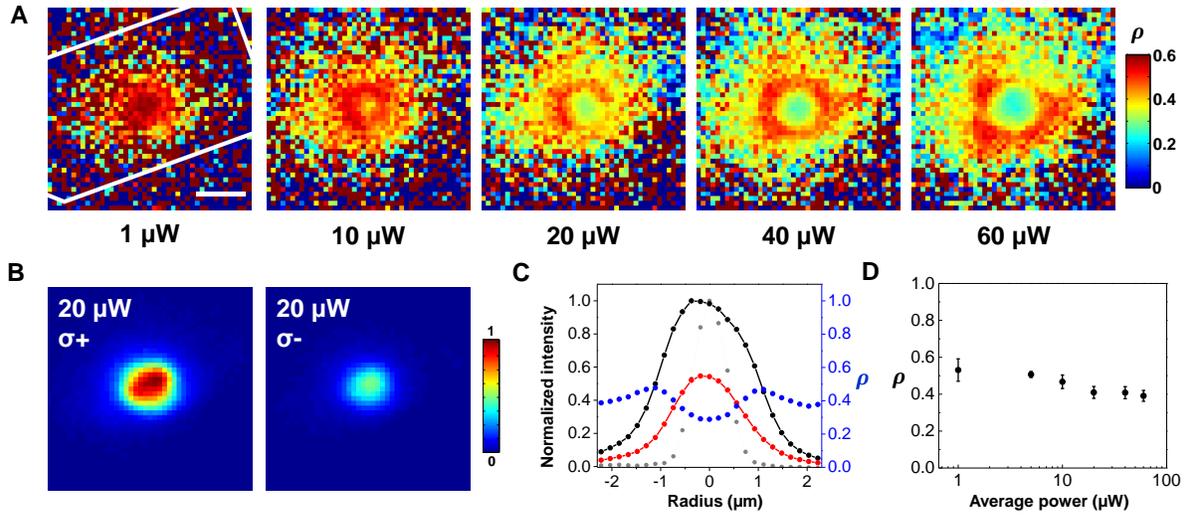

**Figure 3 | Drift-diffusion of valley-polarized interlayer exciton gas.** All plots are for $\sigma^+$ pulsed laser excitation. **(A)** Spatial map of valley polarization under 1-60 μW excitation. Sample outline is shown in white overlay and scale bar is 2 μm. **(B)** Spatial maps of $\sigma^+$ (left) and $\sigma^-$ (right) interlayer exciton photoluminescence with normalized intensity under 20 μW excitation. **(C)** Polarization-resolved spatial profiles of $\sigma^+$ (black) and $\sigma^-$ (red) components of interlayer exciton photoluminescence under 40 μW excitation. The spatial distribution of valley polarization is shown in blue and the laser excitation profile in gray. Line cuts are radially averaged through the excitation center and curves are added as guides to the eye. **(D)** Power dependence of the valley polarization from the spatially integrated $\sigma^\pm$ PL components.

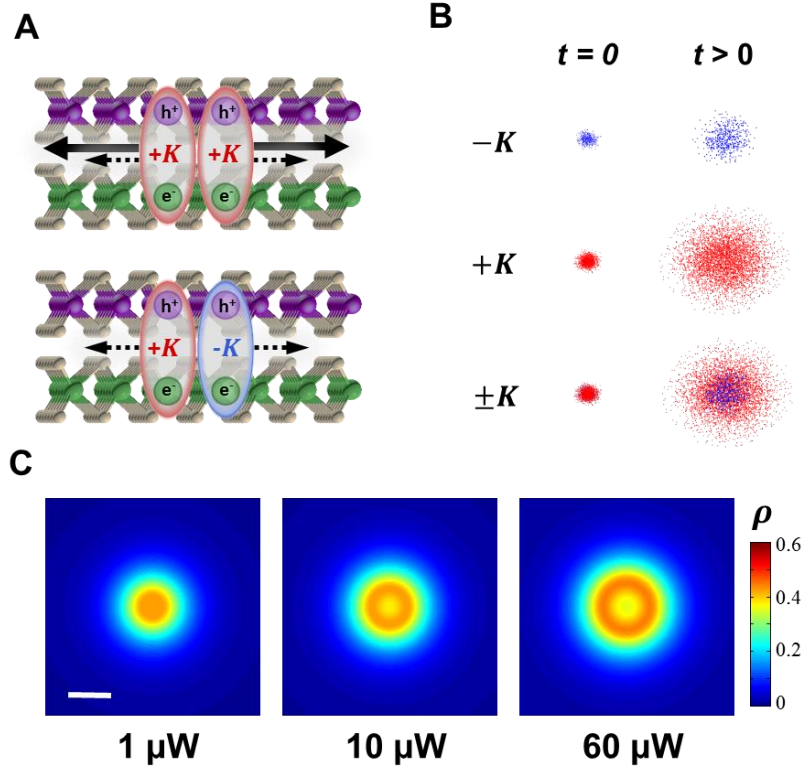

**Figure 4 | Ring formation in valley drift-diffusion. (A)** Schematic of the valley-dependent repulsive interactions between interlayer excitons. Excitons in the same valley (top) experience repulsive forces due to the dipole-dipole interaction (dotted arrows) as well as the exchange interaction (solid arrows), whereas excitons in opposite valleys (bottom) only experience the dipole-dipole interaction. **(B)** Under intense $\sigma^+$ excitation, the majority $+K$ excitons (red) experience larger repulsive forces and thus expand farther than the minority $-K$ excitons (blue), leading to the formation of a ring of valley-polarization in the interlayer exciton gas. **(C)** Simulation of the spatial distribution of interlayer exciton valley polarization at three selected powers. Scale bar is 2 μm.

# Supplementary Materials:

# Expansion of a Valley-Polarized Exciton Cloud in a 2D Heterostructure

Pasqual Rivera[1,†], Kyle L. Seyler[1,†], Hongyi Yu[2], John R. Schaibley[1], Jiaqiang Yan[3,4], David G. Mandrus[3,4,5], Wang Yao[2], Xiaodong Xu[1,6]

Correspondence to: xuxd@uw.edu.

†These authors contributed equally to the work.

## 1. Materials and Methods:

### 1.1 Device Fabrication

WSe$_2$ and MoSe$_2$ monolayers were exfoliated from bulk crystal onto 285 nm SiO$_2$ on n$^+$-doped Si. Linear-polarization-resolved second-harmonic generation (SHG) was used to identify the crystal axes of the WSe$_2$ and MoSe$_2$ monolayers (*1-3*). The intensity of the SHG co-linearly polarized with the excitation laser (Coherent OPA 9800, center ~0.85 eV) was analyzed as a function of crystal orientation. The orientation of the armchair edges was determined by fitting the angular dependence of the intensity to the equation, $I = I_0 \cos^2(3\theta)$, where $I_0$ is the maximum SHG intensity and $\theta$ is the angle between the laser polarization and the armchair edge. We note that this technique is not sensitive to the SHG phase and thus the specific direction along the armchair edge is unknown. Using a polycarbonate-based dry-transfer technique (*4*), we stacked exfoliated monolayers of MoSe$_2$ and WSe$_2$, aligning the armchair edges of the two layers to within 2 degrees. This technique yields heterostructures with twisting angle near 0 or 60 degrees. Standard electron beam lithography was used to pattern contact areas and electron beam evaporation was used to deposit V/Au metal (6/60 nm) contacts. The device in the main text had regions with isolated WSe$_2$ and MoSe$_2$ monolayers, allowing us to confirm the accuracy of stacking process after device fabrication (Fig. S1).

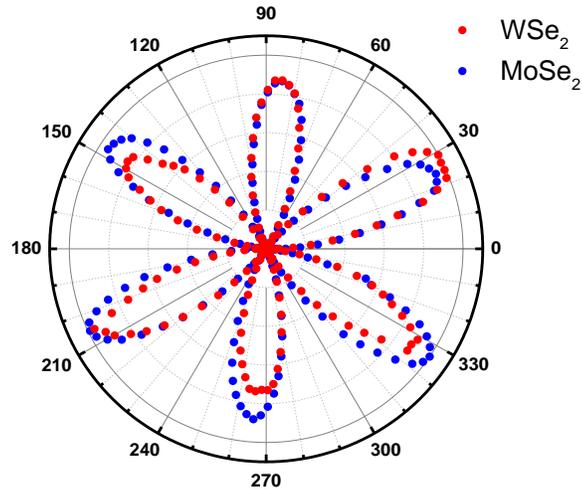

**Figure S1 | Alignment of monolayer crystal axes via second-harmonic generation.** Second-harmonic generation (SHG) intensity parallel to excitation laser linear polarization as a function of crystal angle. The peak SHG intensities correspond to the armchair axes of the monolayer. The spectra are acquired from the individual WSe$_2$ and MoSe$_2$ monolayers after stacking.

### 1.2 Photoluminescence

Photoluminescence (PL) measurements were performed in reflection geometry using a home-built micro-PL setup at 30 K. The PL was collected and analyzed by a grating spectrometer (Andor SR-500i) and charge-coupled device (Andor Newton CCD). For continuous-wave excitation, we employed a power-stabilized and frequency-tunable narrow band (< 50 kHz)

Ti:Sapphire laser (M$^2$ Solstis) at 1.72 eV (720 nm). Pulsed excitation was performed with a supercontinuum laser centered at 1.72 eV with a single-mode Gaussian beam profile focused to 700 nm (FWHM).

For time-resolved PL measurements, the peak wavelength of the interlayer exciton PL was spectrally filtered by the grating spectrometer and measured using a silicon avalanche photodiode (PicoQuant, $\tau$-SPAD) connected to a picosecond event timer (PicoQuant PicoHarp 300) for time-correlated single photon counting. The instrument response of ~400 ps is significantly shorter than the observed interlayer exciton decay times. To accommodate long lifetimes, we used the pulsed excitation source at 100 kHz repetition rate. Data for the polarization decay were plotted after a 10 point re-binning to 2.56 ns time steps in Figs. 2B and S5 and to 1.28 ns in Fig. S6.

For spatially resolved PL measurements, we magnified the PL and set the spectrometer grating to its zero-order mode to image the emission on the CCD. A dichroic beamsplitter and an interference longpass filter ensured that only interlayer exciton PL was collected.

## 2. Supplementary Text and Figures:
## 2.1 Valley-polarized interlayer excitons in supplementary heterostructures.

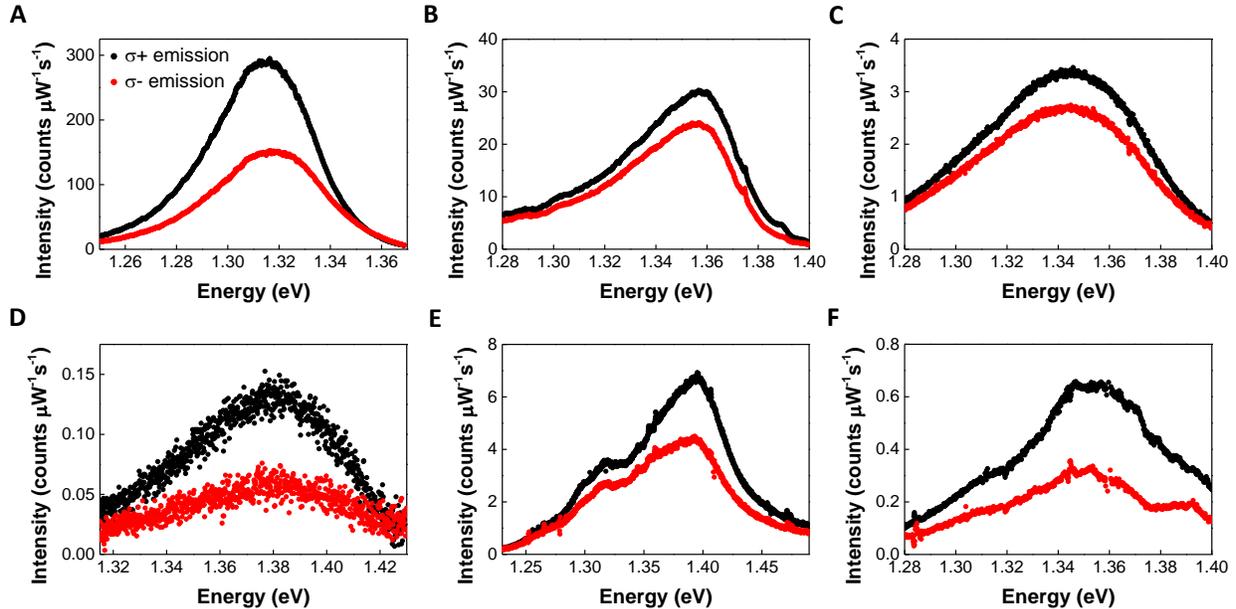

**Figure S2 | Valley-polarized interlayer excitons in supplementary heterostructures.** Circular polarization-resolved interlayer exciton PL spectra from several heterostructures under $\sigma^+$ excitation. Black and red respectively denote $\sigma^+$ and $\sigma^-$ PL emission. The interlayer exciton PL is co-polarized with the helicity of the excitation source for all samples. **(A)** Device presented in the main text with WSe$_2$ on top of MoSe$_2$. **(B-F)** Inverted heterostructures (MoSe$_2$ on WSe$_2$) displaying similar valley polarization. All data are taken at 30K.

## 2.2 Optical pumping of interlayer excitons in heterostructures with AB-like stacking

Here we consider the valley polarization of interlayer excitons in heterostructures with AB-like stacking (twist angle near 60° shown in Fig. S3A). The Brillouin zone of the bottom layer will thus be 180 degree rotation of the top layer (Fig. S3B). In this scenario, the interlayer exciton inherits valley index $+K$ from one layer and $-K$ from the other. Consider the $X_{-,+}$ interlayer exciton as an example. Here the indices denote an electron in $-K_M$ valley of MoSe$_2$ and a hole in $+K_W$ valley of WSe$_2$, as shown in Fig. S3C. It can radiatively recombine through three quantum pathways. The first pathway is through the virtual intermediate state of an intralayer exciton in the $+K_W$ valley: the electron in $-K_M$ conduction band hops to $+K_W$, followed by intralayer recombination in WSe$_2$ and subsequent emission of a $\sigma^+$ photon at the interlayer exciton energy. The second pathway is through a different virtual intermediate state, the intralayer B exciton in the $-K_M$ valley of MoSe2 through the hole interlayer hopping from WSe$_2$ to MoSe$_2$ (Fig. S3C), and the emitted photon is $\sigma^-$ polarized. The radiative recombination can also be facilitated by the interlayer transition dipole between the electron and the hole. The interference of the three pathways leads to elliptically polarized emission of the interlayer exciton (5), with the helicity mainly determined by the relative strength of the electron and hole interlayer hopping rates. When the twisting angle is not exactly 60°, there are three light cones at inequivalent center-of-mass velocities, where the major axes of the elliptical polarization are 120° degree rotation of each other (5). When the PL is from the incoherent sum of the three light cones, the emission can be circularly polarized. Therefore, valley polarization of interlayer excitons in the AB stacking also corresponds to circularly polarized PL.

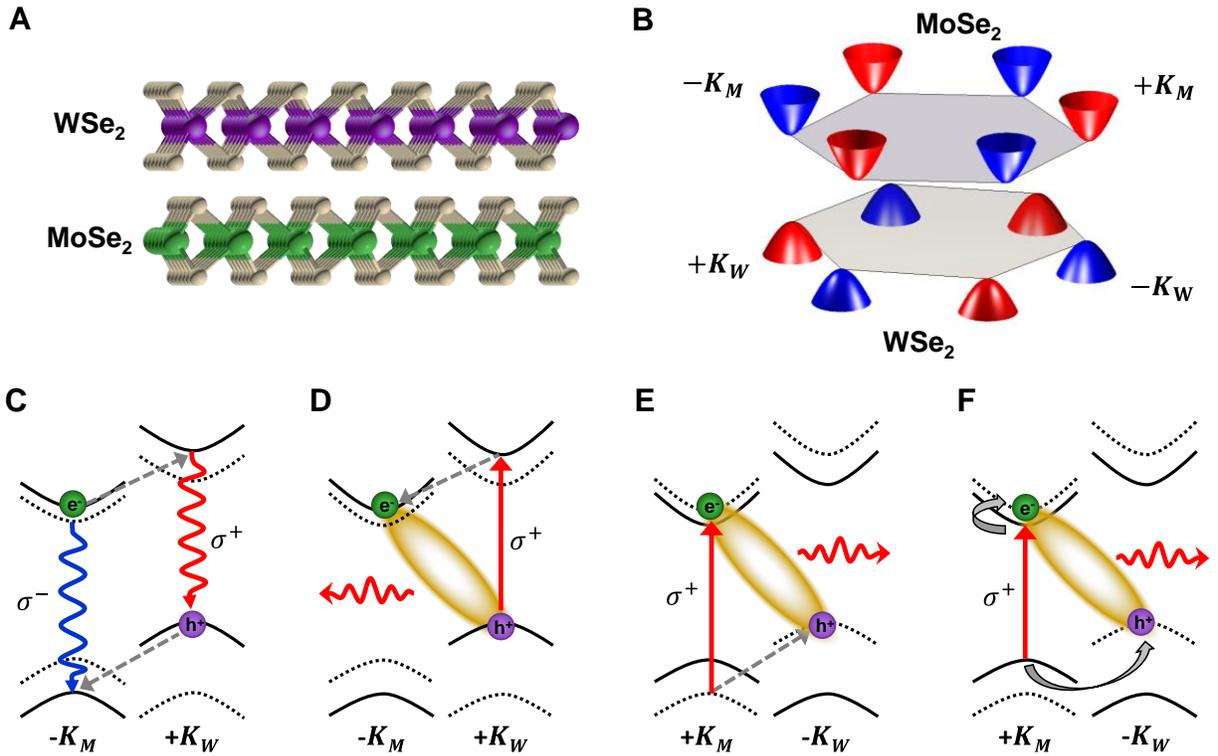

**Figure S3 | Interlayer excitons in heterostructures with AB-like stacking**. (**A**) Side view of AB stacked MoSe$_2$-WSe$_2$ heterostructure. (**B**) Illustration of the band edges in the hexagonal Brillouin zone with twisting angle near 60°. There is 180 degree rotation of the Brillouin zones between the top and bottom layers. (**C**) Illustration of the two quantum pathways contributing to interlayer exciton recombination: the virtual interlayer hopping of an electron (hole) couples to intralayer excitons in WSe$_2$ (MoSe$_2$), emitting $\sigma^+$ ($\sigma^-$) polarized light. Solid (dashed) band indicate spin up (down). (**D**) Optical pumping with $\sigma^+$ polarized light excites the WSe$_2$ A exciton. Spin-conserving interlayer hopping of the electron (gray dashed arrow) forms the interlayer exciton with electron in $-K_M$ valley of MoSe$_2$ and hole in the $+K$ valley of WSe$_2$. (**E**) Optical pumping with $\sigma^+$ polarized light excites the MoSe$_2$ B exciton. Spin-conserving interlayer hopping of the hole (gray dashed arrow) forms the interlayer exciton with electron in $+K$ valley of MoSe$_2$ and hole in the $-K$ valley of WSe$_2$. (**F**) Optical pumping with $\sigma^+$ polarized light excites the MoSe$_2$ A exciton. Interlayer hopping requires the hole to change either its spin or valley index. The formation of interlayer exciton through this channel is expected to be inefficient.

With the lack of information on the relative efficiency of the electron and hole interlayer hopping, the absolute emission helicity of a given interlayer exciton valley configuration (say, $X_{-,+}$) is undetermined. Nevertheless, a general argument leads to the expectation that the emission PL has the same helicity as the circularly polarized excitation. This is because the two recombination pathways resulting in co-polarized emission directly correspond to the two most efficient formation channels of bright interlayer excitons by optical pumping (Figs. S3D and S3E). Other interlayer exciton formation channels, such as that shown in Fig. S3F, require multiple spin and valley flips of carriers, which are not favored. If the electron interlayer hopping is faster than that of the hole, the formation channel via the A exciton in $WSe_2$ is more efficient. In such a case, under $\sigma^+$ excitation, the interlayer exciton will be polarized in the $X_{-,+}$ valley configuration, from which the PL is expected to be $\sigma^+$ polarized. On the other hand, if the formation channels via the B exciton in $MoSe_2$ are more efficient, $\sigma^+$ excitation will generate the interlayer exciton polarized in the $X_{+,-}$ valley configuration, from which the PL is also $\sigma^+$ polarized. If there is AB-like stacking for the samples we have measured (see Fig. S2), the emitted PL is always co-polarized with the incident light, consistent with the expectation.

## 2.3 Gate dependence of interlayer exciton valley polarization

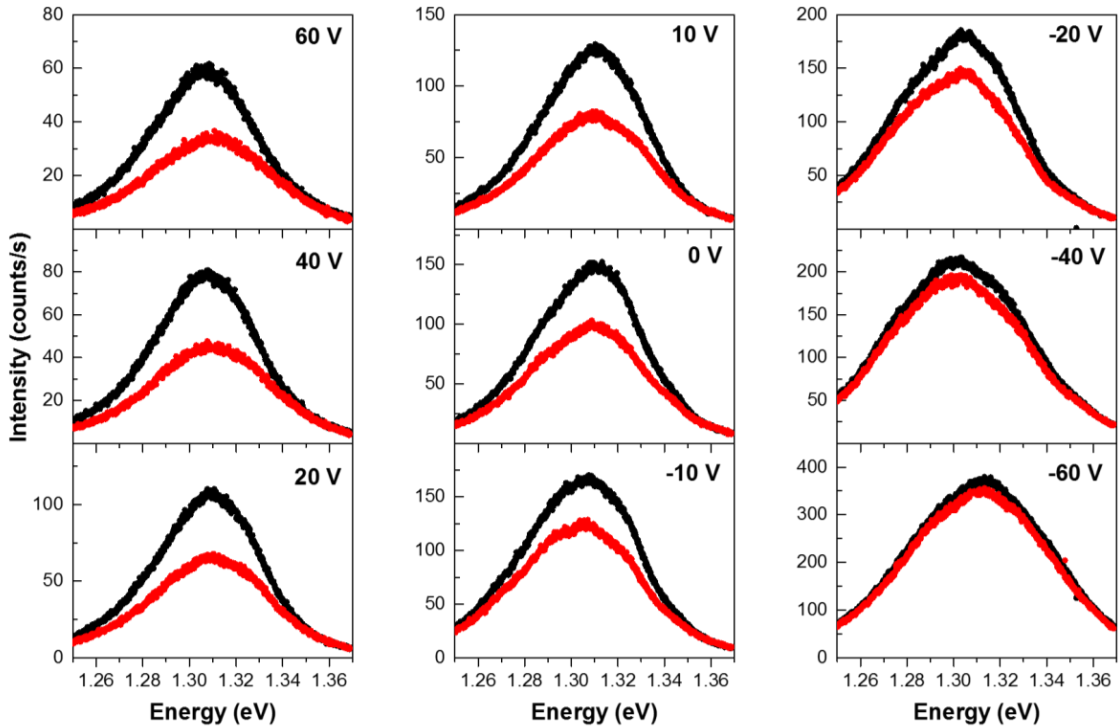

**Figure S4 | Gate dependence of interlayer exciton valley polarization.** Circular polarization-resolved interlayer exciton PL spectra at various gate voltages for the device presented in the main text. The pulsed laser excitation (5 MHz repetition rate) is $\sigma^+$ polarized at 1.72 eV and the co-polarized (black) and cross-polarized (red) components of interlayer exciton PL are detected separately. The degree of valley polarization is largest at high positive gate voltage and becomes nearly negligible at high negative gate voltage.

## 2.4 Gate dependence of polarization dynamics

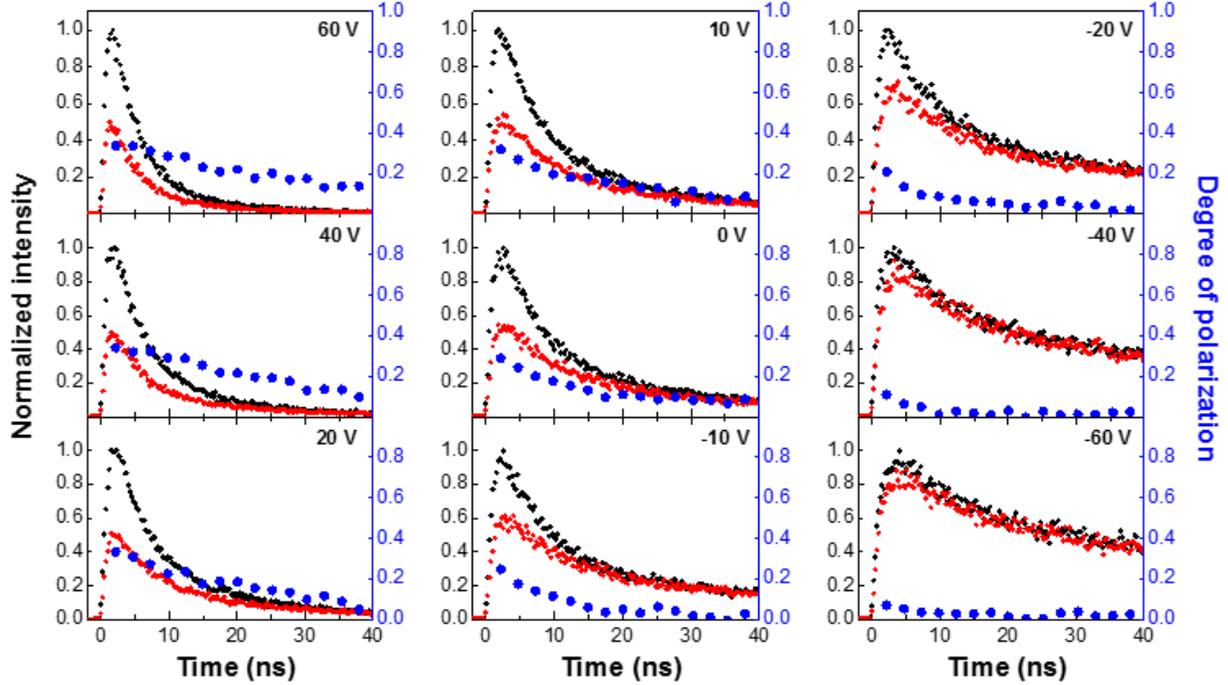

**Figure S5 | Gate dependence of polarization dynamics.** Time and circular polarization-resolved interlayer exciton PL at various gate voltages under pulsed (100 kHz) $\sigma^+$ polarized laser excitation at 1.72 eV for the device in the main text. Co- and cross-polarized emission are shown in black and red, respectively, and the degree of valley polarization is shown in blue. The valley lifetime is longest for large positive gate voltage and decreases with lower gate voltages. At high negative gate voltage, the valley polarization becomes nearly negligible.

## 2.5 Valley lifetime in supplementary heterostructures

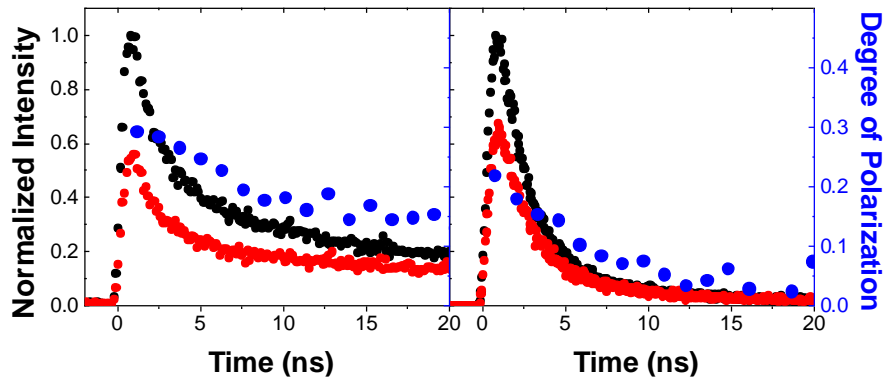

**Figure S6| Valley lifetime in supplementary heterostructures.** Time-resolved interlayer exciton PL from two supplementary heterostructures under $\sigma^+$ polarized pulsed laser excitation at 1.72 eV. Co-polarized and cross-polarized components of interlayer exciton PL are shown in black and red, respectively, and the degree of valley polarization is shown in blue. The stacking order for these devices is inverted (MoSe$_2$ on WSe$_2$) compared to the heterostructure in the main text.

## 2.6 Spatially resolved valley drift-diffusion

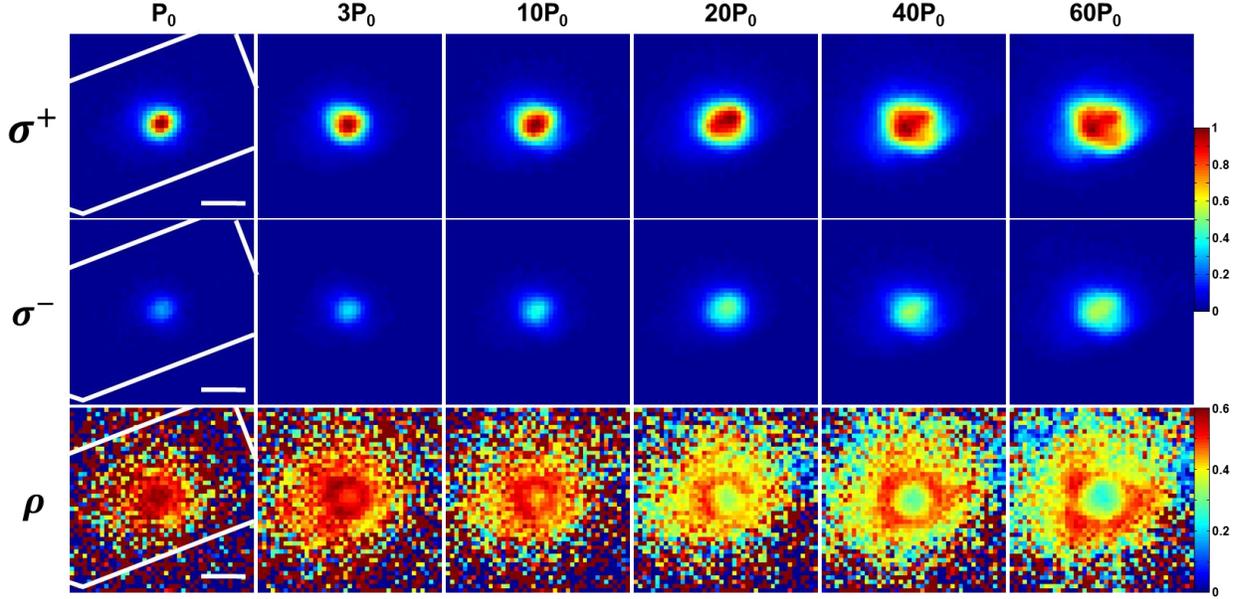

**Figure S7| Spatially resolved valley drift-diffusion.** Spatial maps of valley polarization under $\sigma^+$ pulsed laser excitation at 1.72 eV with $P_0 = 1$ μW. Scale bar is 2 μm. At each power, the spatial profile of co- and cross-polarized emission is shown (normalized to the peak co-polarized intensity) in the top and middle panels, respectively, and the degree of valley polarization is shown in the bottom. The spatial pattern of valley polarization displays the evolution of a ring with increasing diameter under higher excitation power. We note that the maps of the exciton population show noticeable distortion for $P > 20\, P_0$, which indicates the strong role of exciton-exciton interactions with increasing power.

## 2.7 Exchange and dipole interactions between interlayer excitons

Following Reference (*5*), an interlayer exciton as the energy eigenstates of the electron-hole Coulomb interaction in the heterostructure is,

$$X_{\tau'\tau,\mathbf{Q}}(\mathbf{r}_e,\mathbf{r}_h) = \sum_{\Delta \mathbf{Q}} \Phi_I(\Delta \mathbf{Q}) \psi_{\tau',\frac{m_e}{M_0}\mathbf{Q}+\Delta\mathbf{Q},c}(\mathbf{r}_e) \psi^*_{\tau,-\frac{m_h}{M_0}\mathbf{Q}+\Delta\mathbf{Q},v}(\mathbf{r}_h)$$

$$\cong e^{i\mathbf{Q}\cdot\left(\frac{m_e}{M_0}\mathbf{r}_e+\frac{m_h}{M_0}\mathbf{r}_h\right)} \Phi_I(\mathbf{r}_e - \mathbf{r}_h) u_{\tau',0,c}(\mathbf{r}_e) u^*_{\tau,0,v}(\mathbf{r}_h)$$

where **Q** is the kinematic momentum (*5*). $\psi_{\tau',\Delta\mathbf{Q},c}(\mathbf{r}_e) = e^{i(\tau'\mathbf{K}'+\Delta\mathbf{Q})\cdot\mathbf{r}_e} u_{\tau',\Delta\mathbf{Q},c}(\mathbf{r}_e)$ ($\psi^*_{\tau,\Delta\mathbf{Q},v}(\mathbf{r}_h) = e^{-i(\tau\mathbf{K}+\Delta\mathbf{Q})\cdot\mathbf{r}_h} u^*_{\tau,-\Delta\mathbf{Q},v}(\mathbf{r}_h)$) corresponds the $\tau'\mathbf{K}'$ ($\tau\mathbf{K}$) valley electron (hole) Bloch function with $xy$-plane real space coordinate $\mathbf{r}_e$ ($\mathbf{r}_h$), $\Phi_I(\Delta\mathbf{Q})$ describes the **k**-space electron-hole relative motion and $\Phi_I(\mathbf{r}) \equiv \sum_{\Delta\mathbf{Q}} \Phi_I(\Delta\mathbf{Q}) e^{i\Delta\mathbf{Q}\cdot\mathbf{r}}$. We have used the envelope approximation $u_{\tau',\Delta\mathbf{Q},c} \approx u_{\tau',\Delta\mathbf{Q}=0,c}$ and $u_{\tau,\Delta\mathbf{Q},v} \approx u_{\tau,\Delta\mathbf{Q}=0,v}$. We assume a close to $0°$ twisting between the two layers so $X_{++}$ and $X_{--}$ are bright while $X_{+-}$ and $X_{-+}$ are dark (*5*).

Consider a wavepacket of interlayer exciton whose width $W$ is larger than the interlayer exciton Bohr radius $a_B$ but much smaller than the heterostructure sample size $\sqrt{A}$. This wavepacket then has the wavefunction,

$$X_{\tau'\tau,\mathbf{R}}(\mathbf{r}_e,\mathbf{r}_h) \equiv \sum_{\mathbf{Q}} w_{\mathbf{R}}(\mathbf{Q}) X_{\tau'\tau,\mathbf{Q}}(\mathbf{r}_e,\mathbf{r}_h)$$
$$\cong w\left(\frac{m_e}{M_0}\mathbf{r}_e + \frac{m_h}{M_0}\mathbf{r}_h - \mathbf{R}\right)\Phi_I(\mathbf{r}_e - \mathbf{r}_h)u_{\tau',0,c}(\mathbf{r}_e)u^*_{\tau,0,v}(\mathbf{r}_h). \quad (1)$$

Here $w(\mathbf{r} - \mathbf{R}) \equiv \sum_{\mathbf{Q}} w_{\mathbf{R}}(\mathbf{Q})e^{i\mathbf{Q}\cdot\mathbf{r}}$ describes a real space wave packet centered at $\mathbf{R}$. One can define the local density that corresponds to the exciton number in an area comparable to or larger than the wave packet size.

The interaction between two interlayer excitons is illustrated in Fig. S8A, where the solid (dashed) double arrows denote Coulomb attraction (repulsion) between an electron and a hole (two electrons or two holes). We denote the electron (hole) constituent of the $n$-th exciton as $e_n$ ($h_n$) with real space coordinate $\mathbf{d} + \mathbf{r}_{e,n}$ ($\mathbf{r}_{h,n}$) and valley index $\tau'_n$ ($\tau_n$), where $\mathbf{r}_{e,n}$ ($\mathbf{r}_{h,n}$) lies in the $xy$-plane while $\mathbf{d} \equiv d\hat{\mathbf{z}}$ corresponds to the interlayer separation. The exciton-exciton interaction then is

$$\hat{V}_{XX} = V(\mathbf{r}_{e1} - \mathbf{r}_{e2}) + V(\mathbf{r}_{h1} - \mathbf{r}_{h2}) - V(\mathbf{d} + \mathbf{r}_{e1} - \mathbf{r}_{h2}) - V(\mathbf{d} + \mathbf{r}_{e2} - \mathbf{r}_{h1}),$$

with $V(\mathbf{r})$ being the real space form of the Coulomb interaction.

The matrix elements of $\hat{V}_{XX}$ in the two-exciton basis can be separated into the direct and exchange interaction parts. Figure S8B illustrates the direct interaction whose value is

$$V_{dd}(\mathbf{R}_1 - \mathbf{R}_2)$$
$$\equiv \langle X_{\tau'_2\tau_2,\mathbf{R}_2}(\mathbf{r}_{e2},\mathbf{r}_{h2})|\langle X_{\tau'_1\tau_1,\mathbf{R}_1}(\mathbf{r}_{e1},\mathbf{r}_{h1})|\hat{V}_{XX}|X_{\tau'_1\tau_1,\mathbf{R}_1}(\mathbf{r}_{e1},\mathbf{r}_{h1})\rangle|X_{\tau'_2\tau_2,\mathbf{R}_2}(\mathbf{r}_{e2},\mathbf{r}_{h2})\rangle$$
$$= \int |u_{\tau'_1,0,c}(\mathbf{r}_{e1})u^*_{\tau_1,0,v}(\mathbf{r}_{h1})u_{\tau'_2,0,c}(\mathbf{r}_{e2})u^*_{\tau_2,0,v}(\mathbf{r}_{h2})|^2$$
$$\times \left|w\left(\frac{m_e}{M_0}\mathbf{r}_{e1} + \frac{m_h}{M_0}\mathbf{r}_{h1} - \mathbf{R}_1\right)w\left(\frac{m_e}{M_0}\mathbf{r}_{e2} + \frac{m_h}{M_0}\mathbf{r}_{h2} - \mathbf{R}_2\right)\Phi_I(\mathbf{r}_{e1} - \mathbf{r}_{h1})\Phi_I(\mathbf{r}_{e2} - \mathbf{r}_{h2})\right|^2$$
$$\times \left(V(\mathbf{r}_{e1} - \mathbf{r}_{e2}) + V(\mathbf{r}_{h1} - \mathbf{r}_{h2}) - V(\mathbf{d} + \mathbf{r}_{e1} - \mathbf{r}_{h2}) - V(\mathbf{d} + \mathbf{r}_{e2} - \mathbf{r}_{h1})\right)$$
$$\times d\mathbf{r}_{e1}d\mathbf{r}_{e2}d\mathbf{r}_{h1}d\mathbf{r}_{h2}. \quad (2)$$

Since $W > a_B \gg a$ (the lattice constant), $w(\mathbf{r} - \mathbf{R})$ and $\Phi_I(\mathbf{r})$ vary slowly with $\mathbf{r}$ compared to $u_{\tau,0,c/v}(\mathbf{r})$. Here and below we approximate $|u_{\tau,0,c/v}(\mathbf{r})|^2$ by its mean value $\int |u_{\tau,0,c/v}(\mathbf{r})|^2 d\mathbf{r} = 1$, and the periodic part of Bloch function in the integral can be dropped. When the two wave packets are well separated ($|\mathbf{R}_1 - \mathbf{R}_2| \gg W$), we get $\mathbf{r}_{e1},\mathbf{r}_{h1}\sim\mathbf{R}_1$ and $\mathbf{r}_{e2},\mathbf{r}_{h2}\sim\mathbf{R}_2$, so $V_{dd}\sim V(\mathbf{R}_1 - \mathbf{R}_2) - V(\mathbf{d} + \mathbf{R}_1 - \mathbf{R}_2)$ which is long-ranged. For finite $\mathbf{d}$, $V_{dd}$ corresponds to the dipole-dipole interaction between two interlayer excitons.

Figure S8C illustrates the exchange interaction for two excitons with the same valley indices ($\tau'_1 = \tau'_2 = \tau'$ and $\tau_1 = \tau_2 = \tau$):

$$V_{ex}(\mathbf{R}_1 - \mathbf{R}_2)$$
$$\equiv -\langle X_{\tau'\tau,\mathbf{R}_2}(\mathbf{r}_{e2},\mathbf{r}_{h1})|\langle X_{\tau'\tau,\mathbf{R}_1}(\mathbf{r}_{e1},\mathbf{r}_{h2})|\hat{V}_{XX}|X_{\tau'\tau,\mathbf{R}_1}(\mathbf{r}_{e1},\mathbf{r}_{h1})\rangle|X_{\tau'\tau,\mathbf{R}_2}(\mathbf{r}_{e2},\mathbf{r}_{h2})\rangle$$
$$= -\int \Phi_I(\mathbf{r}_{e1} - \mathbf{r}_{h1})\Phi_I(\mathbf{r}_{e2} - \mathbf{r}_{h2})\Phi^*_I(\mathbf{r}_{e1} - \mathbf{r}_{h2})\Phi^*_I(\mathbf{r}_{e2} - \mathbf{r}_{h1})$$
$$\times w^*\left(\frac{m_e}{M_0}\mathbf{r}_{e2} + \frac{m_h}{M_0}\mathbf{r}_{h1} - \mathbf{R}_2\right)w^*\left(\frac{m_e}{M_0}\mathbf{r}_{e1} + \frac{m_h}{M_0}\mathbf{r}_{h2} - \mathbf{R}_1\right)$$
$$\times w\left(\frac{m_e}{M_0}\mathbf{r}_{e1} + \frac{m_h}{M_0}\mathbf{r}_{h1} - \mathbf{R}_1\right)w\left(\frac{m_e}{M_0}\mathbf{r}_{e2} + \frac{m_h}{M_0}\mathbf{r}_{h2} - \mathbf{R}_2\right)$$
$$\times \left(V(\mathbf{r}_{e1} - \mathbf{r}_{e2}) + V(\mathbf{r}_{h1} - \mathbf{r}_{h2}) - V(\mathbf{d} + \mathbf{r}_{e1} - \mathbf{r}_{h2}) - V(\mathbf{d} + \mathbf{r}_{e2} - \mathbf{r}_{h1})\right)$$

$$\times d\mathbf{r}_{e1} d\mathbf{r}_{e2} d\mathbf{r}_{h1} d\mathbf{r}_{h2}. \tag{3}$$

When $|\mathbf{R}_1 - \mathbf{R}_2| \gg W$, $V_{ex} \sim \left(w^*\left(\frac{m_h}{M_0}(\mathbf{R}_1 - \mathbf{R}_2)\right)\right)^2 \left(V(\mathbf{d} + \mathbf{R}_1 - \mathbf{R}_2) - V(\mathbf{R}_1 - \mathbf{R}_2)\right)$ which is exponentially small, so $V_{ex}$ is short-ranged. From the indistinguishability of two electrons (holes), the two-exciton state $X_{\tau'\tau} \otimes X_{\tau'\tau}$ should be written as

$$\frac{1}{2}\Big(X_{\tau'\tau,\mathbf{R}_1}(\mathbf{r}_{e1},\mathbf{r}_{h1})X_{\tau'\tau,\mathbf{R}_2}(\mathbf{r}_{e2},\mathbf{r}_{h2}) - X_{\tau'\tau,\mathbf{R}_1}(\mathbf{r}_{e1},\mathbf{r}_{h2})X_{\tau'\tau,\mathbf{R}_2}(\mathbf{r}_{e2},\mathbf{r}_{h1})$$
$$- X_{\tau'\tau,\mathbf{R}_1}(\mathbf{r}_{e2},\mathbf{r}_{h1})X_{\tau'\tau,\mathbf{R}_2}(\mathbf{r}_{e1},\mathbf{r}_{h2}) + X_{\tau'\tau,\mathbf{R}_1}(\mathbf{r}_{e2},\mathbf{r}_{h2})X_{\tau'\tau,\mathbf{R}_2}(\mathbf{r}_{e1},\mathbf{r}_{h1})\Big)$$

The value $V_{ex}$ then corresponds to a diagonal energy shift of the two-exciton states originating from the Fermion exchange, in addition to the dipole-dipole energy $V_{dd}$ (6-9). $V_{ex}$ is present between two excitons in the same valley while $V_{dd}$ is valley independent.

For an estimation of the strength of $V_{ex}$ and $V_{dd}$ we follow the treatment in Ref. (9), and use the mean field approximation. For a heterostructure sample with spatially inhomogeneous $X_{++}$ density $N_+(\mathbf{r})$ and $X_{--}$ density $N_-(\mathbf{r})$, an $X_{++}$ ($X_{--}$) exciton wave packet at position $\mathbf{R}$ feels a mean potential

$$\Delta E_\pm(\mathbf{R}) = \int V_{ex}(\mathbf{R}-\mathbf{r}) N_\pm(\mathbf{r}) d\mathbf{r} + \int V_{dd}(\mathbf{R}-\mathbf{r})\big(N_+(\mathbf{r}) + N_-(\mathbf{r})\big) d\mathbf{r}$$
$$= N_\pm(\mathbf{R}) \int V_{ex}(\mathbf{R}-\mathbf{r}) d\mathbf{r} + \big(\tilde{N}_+(\mathbf{R}) + \tilde{N}_-(\mathbf{R})\big) \int V_{dd}(\mathbf{R}-\mathbf{r}) d\mathbf{r}$$
$$= N_\pm(\mathbf{R}) \bar{V}_{ex} + \big(\tilde{N}_+(\mathbf{R}) + \tilde{N}_-(\mathbf{R})\big) \bar{V}_{dd}.$$

In the second step above we have used the short-ranged nature of $V_{ex}$, so $\int V_{ex}(\mathbf{R}-\mathbf{r})N_\pm(\mathbf{r})d\mathbf{r} = N_\pm(\mathbf{R}) \int V_{ex}(\mathbf{R}-\mathbf{r}) d\mathbf{r}$. Considering the long-ranged nature of $V_{dd}$ we define $\tilde{N}_\pm(\mathbf{R}) \equiv \frac{\int V_{dd}(\mathbf{R}-\mathbf{r})N_\pm(\mathbf{r})d\mathbf{r}}{\int V_{dd}(\mathbf{R}-\mathbf{r})d\mathbf{r}}$ which is the weighted average of $N_\pm(\mathbf{r})$ over $V_{dd}(\mathbf{R}-\mathbf{r})$. In the experiment $N_\pm(\mathbf{R})$ varies in a length scale of ~μm, so we expect $\tilde{N}_\pm(\mathbf{R}) \approx N_\pm(\mathbf{R})$. In the last line, $\bar{V}_{ex/dd} \equiv \int V_{ex/dd}(\mathbf{R}-\mathbf{r})d\mathbf{r}$, which corresponds to the interaction between two excitons in plane wave states.

Using the hydrogen model ($V(\mathbf{r}) = e^2/\varepsilon r$) for interlayer exciton in coupled quantum wells with homogeneous $\varepsilon$, reference (6-9) have calculated $\bar{V}_{ex/dd}$. The dipole-dipole interaction energy $\bar{V}_{dd} = de^2/\varepsilon$, which increases linearly with $d$. The exchange interaction $\bar{V}_{ex} \approx \left(1 - \frac{d}{0.66 a_0}\right) 6 a_0 e^2/\varepsilon$ also has an approximately linear dependence on $d$, where $a_0$ is the 2D hydrogen model Bohr radius of the intralayer exciton. For small $d$, $\bar{V}_{ex}$ is positive which means the exchange interaction is repulsive for two excitons in the same valley. $\bar{V}_{ex}$ decreases about linearly with $d$ and crosses zero at $d \approx 0.66 a_0$.

It is reasonable to assume the Bohr radius and binding energy of interlayer exciton have same order of magnitudes to those of intralayer excitons in monolayers. Then we can do an order of magnitude estimation of $\bar{V}_{ex}$ and $\bar{V}_{dd}$. The intralayer exciton Bohr radius $a_0 \sim 1 - 3$ nm, while the interlayer distance is $d \sim 0.7$ nm. So we expect $\bar{V}_{ex} \sim a_0 e^2/\varepsilon \sim a_0^2 E_b > 0$, and $\bar{V}_{dd} \lesssim de^2/\varepsilon \sim d a_0 E_b$. Here $E_b \sim e^2/\varepsilon a_0 \sim 0.5$ eV is the intralayer exciton binding energy.

In the layered structure of transition metal dichalcogenides, however, the dielectric constant $\varepsilon$ is anisotropic ($\varepsilon$ in the in-plane direction is much larger than that of the out-of-plane direction), so a simple 2D hydrogen model is not accurate (*10*). Since $\varepsilon$ in the out-of-plane direction is small, $V(\mathbf{d} + \mathbf{r}_e - \mathbf{r}_h)$ should have a slower decay with $d$ compared to the hydrogen model that uses an isotropic $\varepsilon$. This has opposite effects on $\bar{V}_{dd}$ and $\bar{V}_{ex}$. As shown in the Eq. (2), in $\bar{V}_{dd}$, the electron-hole interaction $V(\mathbf{d} + \mathbf{r}_e - \mathbf{r}_h)$ partially cancels the electron-electron and hole-hole interaction energy. While for small $d$ where $\bar{V}_{ex} > 0$, the repulsive exchange interaction $\bar{V}_{ex}$ mainly originates from the electron-hole interaction $V(\mathbf{d} + \mathbf{r}_e - \mathbf{r}_h)$ as shown in Eq. (3). Therefore, at given interlayer separation $d$ (0.7 nm), the hydrogen model overestimates $\bar{V}_{dd}$ and underestimates $\bar{V}_{ex}$.

Finally, we note that the Coulomb exchange interaction also results in another process, as shown in Figure S8D: when the two bright interlayer excitons are from different valleys ($X_{++}$ and $X_{--}$), they will be scattered to become a pair of dark excitons $X_{+-}$ and $X_{-+}$. This can be phenomenologically described as population decay of the bright excitons and has negligible effect on the spatial pattern of polarization.

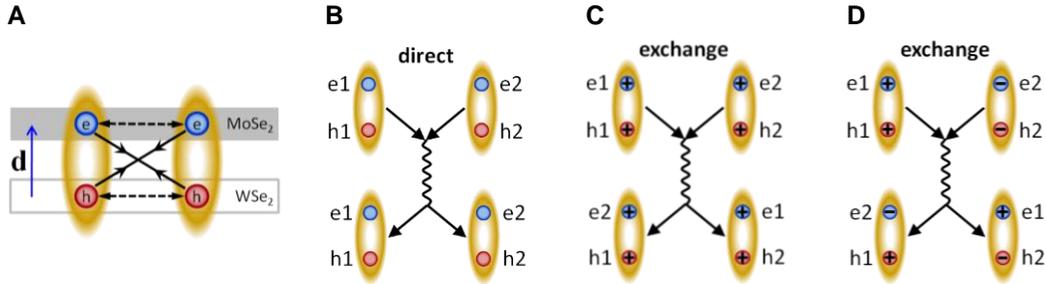

**Figure S8| Interactions between spin-valley polarized interlayer excitons.** (**A**) The interaction between two interlayer excitons. The solid (dashed) double arrows denote the Coulomb attraction (repulsion) between an electron and a hole (two electrons or two holes). (**B**) The diagram of the valley-independent direct interaction between two excitons. (**C**) The exchange interaction between two excitons with the same valley indices. (**D**) The exchange interaction between two excitons with different valley indices. Here, $\pm$ denotes $\pm K$ valleys.

### 2.9 Exciton density under pulsed excitation

Here we estimate an upper bound for the exciton density under pulsed excitation with photon energy of 1.72 eV at 40 µW average power. The optical absorption of the heterostructure is taken as the combined absorption of the individual layers of MoSe$_2$ and WSe$_2$, which are approximately 5% and 10%, respectively (*11*), and we adjust for the observed saturation in PL intensity (Fig. S9). Additionally, we assume the intralayer excitons relax to form interlayer excitons ($X_I$) with unit efficiency. Under pulsed excitation, $\sim 10^5$ photons are absorbed per pulse, forming the same number of interlayer excitons. Due to the short duration of the pulse compared to the exciton lifetime, we normalize the exciton population over the Gaussian profile of the excitation beam (700 nm FWHM), and the peak density of interlayer excitons occurs at the center of the excitation. Immediately after a pulse the interlayer exciton density has a maximum value of $\sim 0.2\, X_I \cdot \text{nm}^{-2}$. Based on the assumptions, this is the upper bound for the interlayer exciton density under our experimental condition.

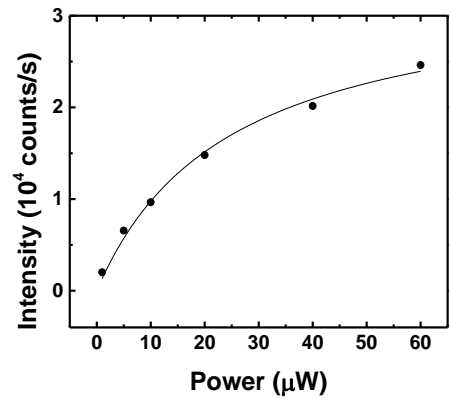

**Figure S9| Power dependence of interlayer exciton PL.** Power dependence of the co-polarized component of interlayer exciton PL intensity (under $\sigma^+$ pulsed excitation at 1.72 eV).

## 2.10 Interlayer exciton drift-diffusion model

Consider a heterostructure system with spatial and time dependent $X_{++}$ density $N_+(\mathbf{r},t)$ and $X_{--}$ density $N_-(\mathbf{r},t)$. The $X_{++}$ exciton at $\mathbf{r}$ feels a mean potential $\bar{V}_{ex}N_+(\mathbf{r},t) + \bar{V}_{dd}(N_+(\mathbf{r},t) + N_-(\mathbf{r},t))$, while the $X_{--}$ exciton at $\mathbf{r}$ feels a mean potential $\bar{V}_{ex}N_-(\mathbf{r},t) + \bar{V}_{dd}(N_+(\mathbf{r},t) + N_-(\mathbf{r},t))$. Such density dependent potential then leads to a gradient force and consequently a drifting velocity $V_{\pm,\text{drift}} = -\alpha_0 \nabla N_\pm - \beta_0 \nabla N_\mp$. Here $\alpha_0 \equiv \frac{\mu}{e}(\bar{V}_{ex} + \bar{V}_{dd}) \sim \frac{\mu}{e}(a_0 + d)a_0 E_b$ and $\beta_0 \equiv \frac{\mu}{e}\bar{V}_{dd} \lesssim \frac{\mu}{e} d a_0 E_b$ (see supplementary discussion 2.8). Here, a mobility value of $\mu \sim 50\, \text{cm}^2 \cdot \text{V}^{-1} \cdot \text{s}^{-1}$ corresponds to $\alpha_0 \sim 10^{-5}\, \mu\text{m}^4 \cdot \text{ns}^{-1}$. For $\beta_0$ currently we only know $\beta_0 < \alpha_0$, but considering the ambiguity of $a_0$ value and the non-hydrogen model correction discussed in section 2.8 of this supplementary text, there could be $\bar{V}_{ex} \gg \bar{V}_{dd}$ and $\alpha_0 \gg \beta_0$. Accordingly, we choose $\alpha_0 = 5\beta_0$ for simulation purposes.

The expansion of the two-valley interlayer excitons under laser generation profile $G_\pm(\mathbf{r},t)$ is described by the drift-diffusion model:

$$\frac{\partial N_+}{\partial t} = G_+(\mathbf{r},t) + D\nabla^2 N_+ + \nabla \cdot \left((\alpha_0 \nabla N_+ + \beta_0 \nabla N_-)N_+\right) - \frac{N_+}{\tau} - \frac{N_+ - N_-}{\tau_v},$$

$$\frac{\partial N_-}{\partial t} = G_-(\mathbf{r},t) + D\nabla^2 N_- + \nabla \cdot \left((\alpha_0 \nabla N_- + \beta_0 \nabla N_+)N_-\right) - \frac{N_-}{\tau} + \frac{N_+ - N_-}{\tau_v}.$$

Here $-(\alpha_0 \nabla N_\pm + \beta_0 \nabla N_\mp)N_\pm$ is the excitonic drift current for the $\pm \mathbf{K}$ valley. $D = \mu k_B T/e \sim 0.01\, \mu\text{m}^2 \cdot \text{ns}^{-1} = 0.1\, \text{cm}^2 \cdot \text{s}^{-1}$ ($T\sim 30$ K) is the diffusion constant, $\tau$ is the exciton life time, and $\tau_v$ is the valley relaxation time. We note that processes such as biexciton formation, Auger scattering, and the population relaxation between bright and dark excitons are not included in this model, which can be oversimplified for a quantitative comparison with experiments. Nevertheless the main qualitative features observed (i.e. the ring-like polarization pattern) can be well reproduced using reasonable parameters, where the valley dependent exchange and the valley independent dipole-dipole repulsive interactions are the main causes.

For the pulsed laser with a very short duration, we write the exciton density distribution just after the pulse as $N_+(\mathbf{r}, t=0) = N_0 e^{-r^2/a^2}$ and $N_-(\mathbf{r}, t=0) = \delta \cdot N_+$, with $N_0$ the peak exciton density and $a$ the laser half-width. Let the dimensionless variable $\rho_\pm(\mathbf{r},t) \equiv N_\pm(\mathbf{r},t)/N_0$, then the evolution is governed by

$$\frac{\partial \rho_+}{\partial t} = D\nabla^2 \rho_+ + \nabla \cdot \left((\alpha \nabla \rho_+ + \beta \nabla \rho_-)\rho_+\right) - \frac{\rho_+}{\tau} - \frac{\rho_+ - \rho_-}{\tau_v},$$

$$\frac{\partial \rho_-}{\partial t} = D\nabla^2 \rho_- + \nabla \cdot \left((\alpha \nabla \rho_- + \beta \nabla \rho_+)\rho_-\right) - \frac{\rho_-}{\tau} + \frac{\rho_+ - \rho_-}{\tau_v}.$$

In the above we have written $\alpha \equiv N_0 \alpha_0$ and $\beta \equiv N_0 \beta_0$. Under a peak density $N_0 \sim 10^{12} - 10^{13}\, \text{cm}^{-2}$, $\alpha \sim 0.1 - 1\, \mu\text{m}^2 \cdot \text{ns}^{-1}$.

## 2.11 Drift-diffusion model simulation results

Here we present simulation results obtained from the drift-diffusion model. Figure S10 shows the results using the model parameters listed in Table S1 and exciton densities $N_0 = [0.0012, 0.005, 0.009, 0.014, 0.020, 0.023]$ nm$^{-2}$. These values are the estimated exciton densities corresponding to excitation powers of 1-60μW, as determined in section 2.8. The simple model captures the qualitative features from the experiment, namely the evolution of an expanding ring pattern under increasing excitation powers.

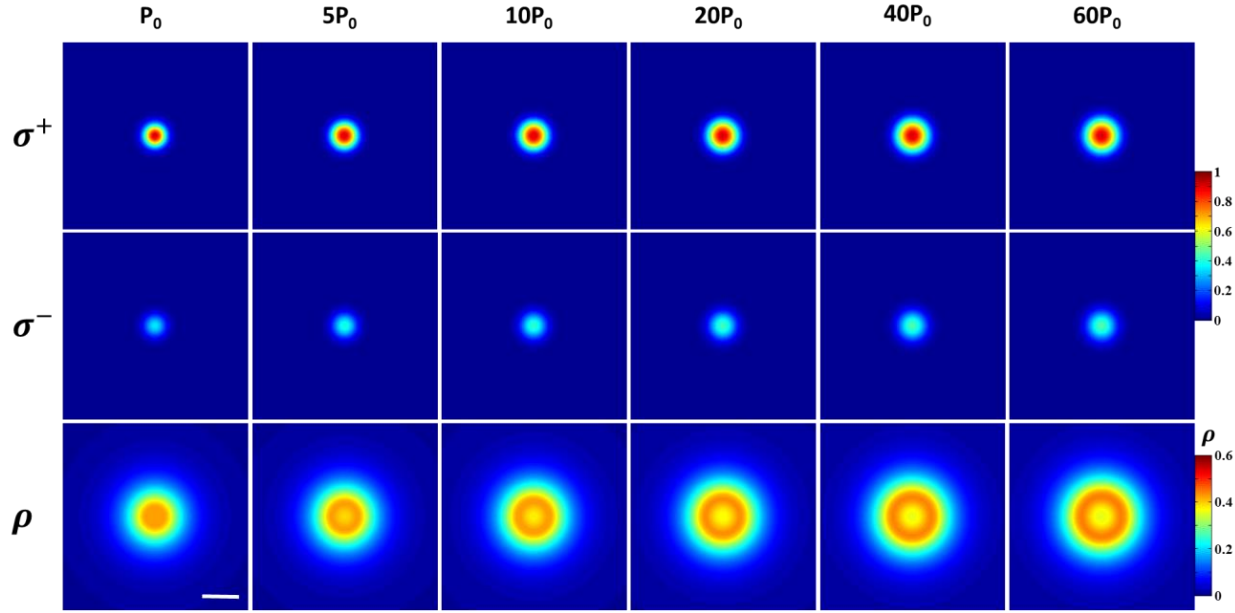

**Figure S10 | Simulated interlayer exciton valley drift-diffusion.** Simulated spatial maps of valley polarization under $\sigma^+$ pulsed laser excitation for exciton densities corresponding to excitation powers of 1-60 μW. At each power, the spatial profile of co- and cross-polarized interlayer exciton density is shown in the top and middle panels (normalized to the peak co-polarized density), respectively, and the degree of valley polarization is shown in the bottom. The spatial pattern of valley polarization displays the evolution of a ring with increasing diameter under higher excitation power. $P_0 = 1$ μW and the scale bar is 2 μm.

| $a$ (nm) | $\tau$ (ns) | $\tau_v$ (ns) | $\alpha_0$ ($10^{-6}$ μm$^4 \cdot$ ns$^{-1}$) | $\beta_0$ ($10^{-6}$ μm$^4 \cdot$ ns$^{-1}$) | $D$ (μm$^2 \cdot$ ns$^{-1}$) | $\delta$ |
|---|---|---|---|---|---|---|
| 420 | 10 | 40 | 10 | 2 | 0.005 | 0.2 |

**Table S1: Drift-diffusion model parameters for CW and pulsed excitation.**